\documentclass{article}

\usepackage{arxiv}

\usepackage[utf8]{inputenc} 
\usepackage[T1]{fontenc}    

\usepackage{lmodern}
\usepackage{enumitem}
\usepackage{amssymb}
\usepackage{wrapfig}
\usepackage{lscape}
\usepackage{rotating}
\usepackage[section]{placeins}
\usepackage[superscript,biblabel]{cite}
\usepackage[hidelinks]{hyperref}       
\usepackage{longtable}
\usepackage{url}                       
\usepackage{booktabs}                  
\usepackage{amsfonts}                  
\usepackage{nicefrac}                  
\usepackage{microtype}                 
\usepackage{lipsum}                    
\usepackage{graphicx}
\usepackage{doi}

\usepackage[export]{adjustbox}
\let\oldincludegraphics\includegraphics
\renewcommand{\includegraphics}[2][]{
  \centering
  \oldincludegraphics[#1,max size=0.9\textwidth]{#2}
  }
\usepackage[width=\textwidth,font=footnotesize]{caption}

\newcommand{\process}[1]{\textsf{\it #1}}
\newcommand{\task}[1]{\textsf{\it #1}}

\title{Software Development Processes in Ocean System Modeling}

\author{ \href{https://orcid.org/0000-0002-5464-8561}{\includegraphics[scale=0.06]{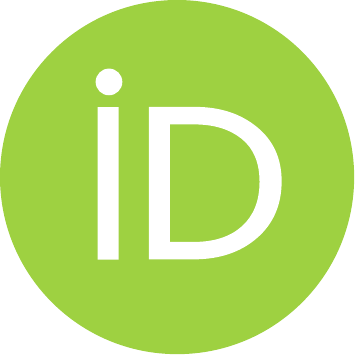}\hspace{1mm}Reiner~Jung}\\
	Software Engineering Group\\
	Department of Computer Science\\
	Kiel University\\
	Kiel, Germany\\
	\texttt{reiner.jung@email.uni-kiel.de} \\
	\And
	\href{https://orcid.org/0000-0003-4060-2754}{\includegraphics[scale=0.06]{orcid.pdf}\hspace{1mm}Sven~Gundlach} \\
	Software Engineering Group\\
	Department of Computer Science\\
	Kiel University\\
	Kiel, Germany\\
	\texttt{sven.gundlach@email.uni-kiel.de} \\
	\AND
	\href{https://orcid.org/0000-0001-6625-4335}{\includegraphics[scale=0.06]{orcid.pdf}\hspace{1mm}Wilhelm~Hasselbring} \\
	Software Engineering Group, Department of Computer Science\\
	Kiel University\\
	Kiel, Germany\\
	\texttt{hasselbring@email.uni-kiel.de} \\
}

\date{August 2021}

\renewcommand{\headeright}{}
\renewcommand{\undertitle}{}
\renewcommand{\shorttitle}{Software Development Processes in Ocean System Modeling}

\hypersetup{
pdftitle={Software Development Processes in Ocean System Modeling},
pdfsubject={cs.SE,cs.CE},
pdfauthor={Reiner~Jung,Sven~Gundlach,Wilhelm Hasselbring},
pdfkeywords={Scientific modeling, software development processes, process modeling, thematic analysis},
}

\begin{document}
\maketitle

\begin{abstract}
Scientific modeling provides mathematical abstractions of real-world systems and builds software as implementations of these mathematical abstractions. Ocean science is a multidisciplinary discipline developing scientific models and simulations as ocean system models that are an essential research asset.

In software engineering and information systems research, modeling is also an essential activity. In particular, business process modeling for business process management and systems engineering is the activity of representing processes of an enterprise, so that the current process may be analyzed, improved, and automated.

In this paper, we employ process modeling for analyzing scientific software development in ocean science to advance the state in engineering of ocean system models and to better understand how ocean system models are developed and maintained in ocean science. We interviewed domain experts in semi-structured interviews, analyzed the results via thematic analysis, and modeled the results via the business process modeling notation BPMN.

The processes modeled as a result describe an aspired state of software development in the domain, which are often not (yet) implemented. This enables existing processes in simulation-based system engineering to be improved with the help of these process models.

\end{abstract}


\newpage

\section{Introduction}

In scientific modeling, mathematical abstractions of real world systems are defined and software is built as implementation of these mathematical abstractions.
The resulting software systems are called \textit{models} of the real world system, which are then executed to simulate the real world.
These models are used to make predictions and to understand emerging properties of the real world. 
Computational science and scientific computing are multidisciplinary fields lying at the intersection of mathematics and statistics, computer science, and core disciplines of science.
This involves the development of scientific models and simulations to understand natural systems answering questions that neither theory nor experiment alone is equipped to answer.\cite{Ruede2018}

In the past, scientific computing and software engineering have separately developed methods and approaches, with limited collaboration.\cite{CiSE2018}
To a great extent, software engineering focused on creating, evaluating techniques, methods and tools to develop and maintain large and complex software systems, mainly in the domains of embedded software systems and information systems.\cite{Randell2018}
However, little of this body of knowledge has been transferred to scientific computing.

An application area of scientific modeling is to build ocean system models. These ocean system models address certain aspects of the ocean, such as CO$_2$ uptake.\cite{wanninkhof_global_2013}
Many such models, which are complex software systems, have been developed and extended over decades by scientists from different disciplines.\cite{ClimateModels2015}
Thus, they are an interdisciplinary effort.
They have grown in size and complexity which affects the effort necessary to improve and change models for addressing new research questions.
Despite the increasing importance of software-based modeling to the scientific discovery process, well-established software engineering practices are rarely adopted in computational science.\cite{CiSE2018}
As a step to advance the state in engineering of developing and maintaining ocean system models, we employed process modeling for analyzing the development and application process of these system models in ocean science.

The best way to gain insights into a domain is observing and asking domain experts, in our case ocean model developers, model users, and research software engineers.
It is the main goal to understand how ocean modelers and engineers create, maintain, extend and use models. Based on this understanding, we later intend to create tools and services that support the research work and improve collaboration in ocean modeling.
We chose semi-structured interviews as the method to address key issues, to understand how modeling is done in ocean science and yet to be flexible enough to adapt if certain new aspects emerge.
We interviewed ocean model developers, scientific modelers and research software engineers who work in various research groups at  universities and large-scale research facilities, such as the GEOMAR Helmholtz Centre for Ocean Research Kiel, the German Climate Computing Center (DKRZ) and the Max Planck Institute for Meteorology (MPI). 

For the analysis of the interviews, we chose the Thematic Analysis~(TA)\cite{Braun2006} approach tailored to our questions, since they directly address processes forming a methodological frame for the analysis.
As our questions are addressing well defined areas, our themes carry semantics, i.e., they are directly expressed in the interviews.

In software engineering and information systems research, modeling is also an essential activity.\cite{jensen_model-based_2009}
Conversely to (ocean) science, in software engineering and in information systems research models are built as abstractions of the software systems and the software development processes themselves.
A software architecture description is an example for such a model in software engineering and in information systems research.\cite{SA2018}
In particular, business process modeling for business process management 
is the activity of representing processes of an enterprise, such that the current process may be analyzed, improved, and automated. 
In this paper, we employ the established Business Process Modeling Notation~(BPMN)\cite{chinosi_bpmn2012} method method of process modeling for analyzing scientific modeling processes.

The resulting modeled processes show an ideal state of the observed software development process that is already achieved in the domain but only partially used or implemented for special applications. 
Thereby existing development and application processes can be reviewed using the analysis technique or can be optimized with the input of these processes.
Our employed research methods are summarized in Section~\ref{sec-research-method}.
The results of the thematic analysis are described in Section~\ref{sec-thematic-analysis}.
The elicited software development processes in ocean science are presented in Section~\ref{sec-resulting-themes}.
Related work is discussed in Section~\ref{sec-related-work}, before Section~\ref{sec-conclusion} draws our conclusions.
\section{Research Methods}\label{sec-research-method}

To study software development processes in ocean science, we conduct semi-structured interviews and analyze the interview results via the Thematic Analysis~\cite{Braun2006} approach.
Section~\ref{sec-semi-structured-expert-interview} reports on our interview setup, before Section~\ref{ssec-thematic-analysis} explains how we used Thematic Analysis for analyzing the interview results.

\subsection{Semi-structured Expert Interviews}\label{sec-semi-structured-expert-interview}

As a qualitative research method, interviews allow to investigate new ideas, concepts and domains.\cite{seidman_interviewing_2019,Kaiser2014}
Interviews can be characterized by the kind of interviewees (e.g., experts), number of interviewees, and how questions are asked ranging from only an initial question (narrative interview) to structured interviews with detailed questionnaires.
The semi-structured interviews, that were conducted in our study, lie in between narrative interviews and fully structured interviews.
We follow an interview guide which comprises of introductory questions, e.g, profiles, followed by the main section with a few central questions and explanations.
These questions and explanations are used to steer the interviews, but still allow enough room to explore the overall topics of the interviews.

Due to the SARS-CoV-2 pandemic, most of the interviews were conducted via video conference.
As we used semi-structured expert interviews, we developed an interview guide addressing the context and background of the interviewees, their work process, the technical environment they work in, and whether they use software engineering methods.

\paragraph{Employed Software Tools}
Thematic Analysis requires complete transcriptions of the interviews.
All of our interviews were recorded in an audio format for later automated speech recognition.
As speech recognition software, a Kaldi Gstreamer Server was used with the latest pretrained model.\cite{milde-koehn-18-german-asr}
The finalization was done by manually proofreading the transcript and analyzing the raw data using the audio software ocenaudio.\cite{ocenaudio}
For the further TA, including familiarization, generation, searching and reviewing, the R package for Qualitative Data Analysis RDQA \cite{RQDA} was employed.
Final processing, including defining and producing codes, was done using a SQL database, customized R-scripts and some Java programs.

\paragraph{Participants}
We conducted all interviews with two interviewers and in most cases one interviewee to reduce the risk of skipping over interesting point.
In total we had 9 interviews with 10 interviewees. 
The interviewees had worked or were actively working at GEOMAR, MPI, DKRZ and Kiel University and have both scientific and technical backgrounds (cf. Section~\ref{sec-roles-and-use-cases}).

\paragraph{Structured Interview Guide}
The interviews were structured into introduction, process, environment and methods, and was directed toward the modeling process.
Each part was used for an initial theme of the TA, namely Interviewee Profile, Software Development Processes, Infrastructure Environment and Software Engineering Methods:
\begin{itemize}
  \item \textbf{Interviewee Profile} The first introductory questions address the background of the scientists and engineers as well as their job characteristics and research interests. This way, we identify the demographics of our participants. This allows to understand the involved disciplines and topics addressed within the domain.
  \item \textbf{Software Development Processes} The second part is directed on how scientists and engineers work in ocean modeling.
Here the main interest is in their work processes and interactions with others.
  \item \textbf{Infrastructure Environment} The next block addresses the software and infrastructure environment used to develop, test, and run their models.
This is important, as the acceptance of new tools and practices depends on the ability to use them with reasonable effort.
  \item \textbf{Software Engineering Methods} The interview closes with questions regarding software engineering methods, such as testing, static code checking, and feature management.
\end{itemize}

\subsection{Thematic Analysis of the Interview Results}\label{ssec-thematic-analysis}

Qualitative research provides a wide area of different methods and methodologies to analyze data and especially interviews.
These methods are usually developed in the social sciences and then transferred to other disciplines, including software engineering. 
For this study, we chose Thematic Analysis~(TA) as the method to analyze our interview data. 
TA has a long history which is widely used in psychology and the social sciences to analyze qualitative data including  images, videos and interviews.
Following the conceptualization of Braun,\cite{Braun2006} it is a flexible method that can be customized towards our research goals.
The central concept of TA are themes.
A \emph{theme} captures an idea, concept or pattern within the data which seems relevant in relation to the research question.
The relevance of a theme is not defined by quantifiable measures, but by the \emph{relevance} with respect to the research goals.

TA can be used to provide a \emph{descriptive} result for the whole data set, e.g., all interview transcripts, or for an \emph{in-depth} analysis to identify specific latent themes.
A \emph{latent theme} in TA is a theme which is not explicitly mentioned in the questions and answers during an interview.
Instead it describes an underlying pattern.
For example, the pattern that there are certain roles in the ocean modeling domain and that there are competing priorities between these roles.
In contrast, a \emph{semantic theme} is closer to the data set and more descriptive.
TA can be applied to any media including interviews and transcripts.
All media are considered \emph{data} and all media used for a specific TA is called \emph{data set}.

In our study, we combined inductive and deductive analysis.
\emph{Inductive analysis} is a bottom up approach where coding is done without a pre-existing coding frame or an analytic preconception.
In an inductive analysis, the underlying paradigm is constructivistic focusing on socio-cultural contexts and structural conditions.
Meanings and experiences of interviewees are seen as socially (re-)produced entities.
This method is mainly useful when analyzing a previously unknown domain, or to avoid influencing the analysis results by personal bias.
\emph{Deductive analysis}, in contrast, is driven by a particular research interest in an area.
Thus, codes and themes are influenced by research questions.
Essentially, an inductive analysis allows to identify properties of a domain that were not previously considered.
This is especially useful when trying to understand new domains.
However, it may produce less specific answers regarding a specific research goal.
Whereas a deductive analysis narrows the focus to specific research interests and provides more in-depth results, it may miss properties that were not considered in advance.
As we aim to understand processes, we started with specific research questions and, therefore, with a deductive analysis but complemented this by a subsequent inductive analysis.

Qualitative data analysis
examines and interprets data to understand what it represents.
The TA method comprises six phases which can be re-iterated multiple times, depending on modification to the themes, codes and text extracts:

\paragraph{Phase 1: Familiarization with the data}
The researcher familiarizes him- or herself by actively reading the whole data set -- at least once -- before coding. Coding is the process of labeling and organizing qualitative data to identify different themes and the relationships between them. 
During familiarization, the researcher collects initial ideas and patterns including the reasons for their relevance.

\paragraph{Phase 2: Generate initial codes}
Based on these ideas, the researchers identify an initial set of codes.
When coding, they assign labels to words or phrases that represent important (and recurring) themes in each response. 
In a data-centric approach the codes rely primarily on the data, while in a theory-centric approach, the codes are influenced by the set of research questions the researchers have towards the data set.
Each code refers to a meaning within a data extract.
However, a text extract may relate to multiple codes, such as the text extract \emph{'we use git to manage code and contributions'} may be coded as both \emph{'Version control with git'} and \emph{'Code management'}.
Thus, this phase produces a set of codes where each code is associated with one or more extracts from the whole data set.

\paragraph{Phase 3: Searching for themes}
Themes represent patterns and relevant topics -- also referred to as narratives -- occurring in or emerging from the interviews.
To create themes, codes are grouped and subsequently associated with potential themes.
Themes can be interrelated.
The relationship includes contrasting views and hierarchy, i.e., a theme to sub-theme relationship.
This results in a \emph{theme map} that contains a set of candidate themes, related to codes and data extracts.
In this phase no code or extract gets discarded.
In case a code does not (yet) match any theme, it can be associated to a theme called \textit{miscellaneous}.

\paragraph{Phase 4: Reviewing and refining themes}
The review and refinement of the thematic map comprises two steps.
First, each theme is analyzed regarding its associated extracts to check whether they form a coherent pattern, i.e., they address the same issue, topic or narrative.
In case an extract does not fit, it is discarded or moved to another theme.
Here, themes can be created, split, merged or removed.
Second, the resulting themes must be reflected based on the whole data set, if they are accurate representations of the data set.
Therefore, it is crucial to revisit the associated text segments and their context. 
In case relevant aspects are missing, additional codes can be added.

\paragraph{Phase 5: Defining and naming themes}
The essence of each theme is identified and an internally consistent account of the theme is compiled.
In case themes are too complex and diverse, they are split and further sub-themes are identified.

\paragraph{Phase 6: Produce the report}
Finally, a report is compiled covering all themes.
The report must cover the complete story of the whole data, provide specific descriptions of all themes and their relationships.

~

Before selecting TA, we considered other analysis approaches and especially Grounded Theory as means to analyze the interview data.
However, as reported by Stol et al.,\cite{Stol2016} Grounded Theory is often not used properly in software engineering research. 
Grounded Theory aims to establish new concepts and ideas based on interview data, without using an explicit research question during the analysis.
Since our semi-structured interviews are based on specific interview questions, we chose TA for our analysis.

\section{Results of the Thematic Analysis for Ocean System Modeling}
\label{sec-thematic-analysis}

We analyzed the interviews utilizing Thematic Analysis~(TA) as described in Section~\ref{ssec-thematic-analysis}.
We identified multiple themes that support a general understanding of the way scientists and engineers work in the domain.
For this paper, we focus our analysis on the software development processes performed by scientists and engineers in ocean system modeling.
We followed the six phase process of TA, as introduced in Section~\ref{ssec-thematic-analysis}.
For the transcription, we relied on speech to text software to support the text generation (cf. Section~\ref{sec-semi-structured-expert-interview}).
Still, the result required intensive manual adjustments due to the specific topic of our study.
We were able to familiarize ourselves with the interview text and audio in great detail in this phase.

Based on the transcripts, we identified initial code candidates and started coding with the RQDA coding tool.\cite{RQDA}
We created codes based on the interview data, following a data-centric thematic analysis approach, and selected codes and categories based on our research questions, following a theory-centric thematic analysis approach.
During the initial coding, we generated new codes for each interview resulting in some duplication, which had to be merged at a later stage.
Based on the coding, we identified processes and role-related themes, as well as other domain-specific themes.
We discovered relationships between these themes and mapped all codes to at least one theme.
Subsequently, we reviewed and refined the identified themes following the approach in Section~\ref{ssec-thematic-analysis}.
Thus, similar codes were merged, codes were moved to more fitting themes and removed if they were out of scope.
Finally, all interviews were re-read reflecting each theme.
During this process new codes and themes emerged, which had to be addressed accordingly.

\begin{figure}[!htb]
	\centering
	\adjustimage{height=0.58\textheight,angle=90}{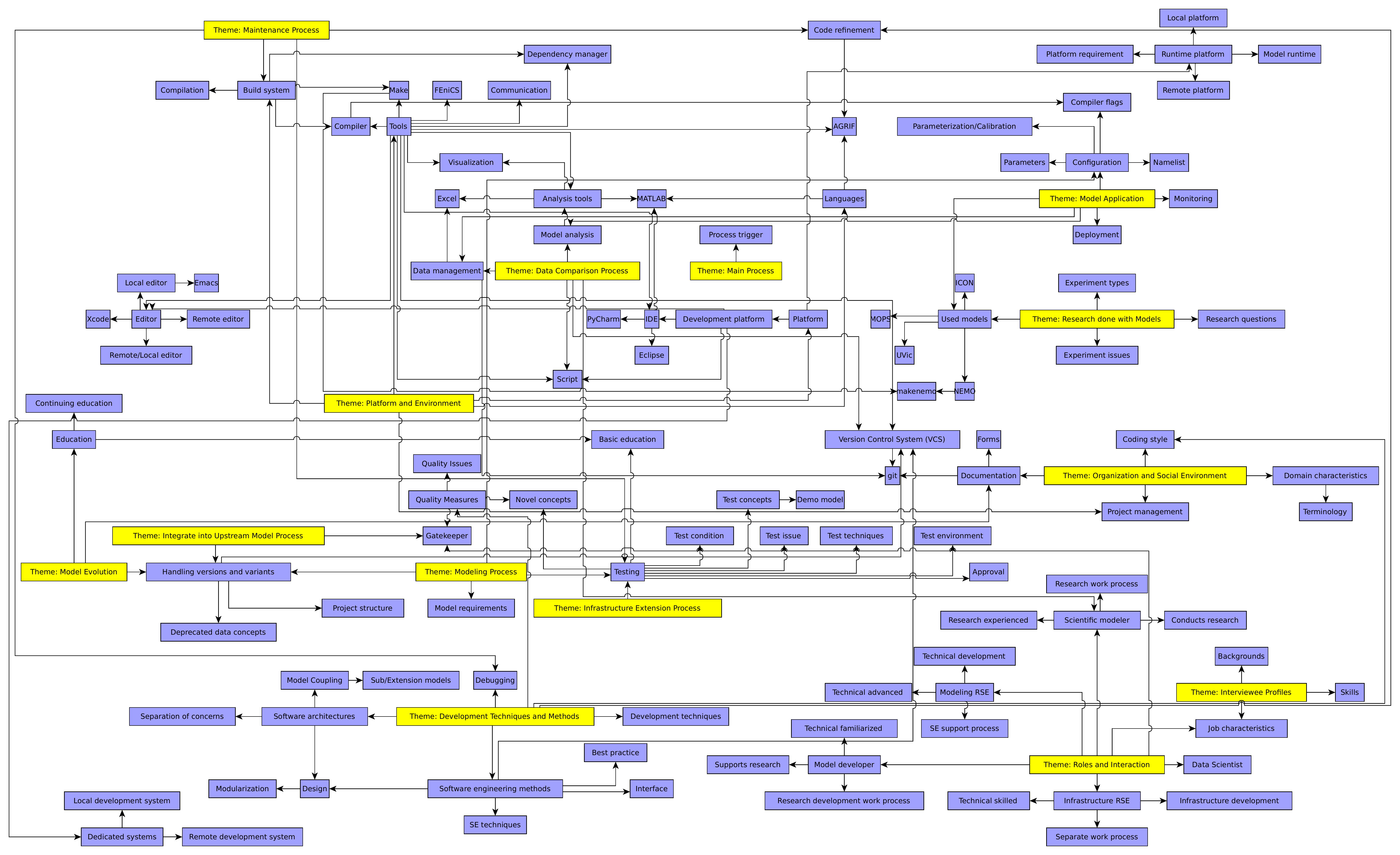}
	\caption{Excerpt of the thematic map depicting the main resulting themes in yellow and sub-themes in blue. Refer to \url{https://oceandsl.uni-kiel.de/graph/} for the complete, interactive map.}
	\label{fig-thematic-map}
\end{figure}

As a result, we obtained a large theme map.
The map depicted in Figure~\ref{fig-thematic-map} shows only the process-centric excerpt of the main themes and associated sub-themes, which are relevant for this paper.
The complete map is provided as supplementary material in the form of on online graph at
\begin{quote}
	\url{https://oceandsl.uni-kiel.de/graph/}
\end{quote}
This online graph allows you to interactively explore our coding results. In the present paper, we only display an extract in Figure~\ref{fig-thematic-map}.
In the following section, we will present the software development processes that were derived from this thematic map.

\section{Elicited Software Development Processes for Ocean System Modeling}\label{sec-resulting-themes}

Based on the thematic analysis, we gained insight into the work processes and roles of scientists and engineers in ocean modeling.
In this section, we model the resulting processes via BPMN.
As activities of these processes are performed by specific roles, we first introduce the core roles associated with these processes in Section~\ref{sec-roles-and-use-cases}, followed by the elicited processes in Section~\ref{sec-main-process} to Section~\ref{sec-infrastructure-process}.

\newcommand{\imgscale}{0.6}

\subsection{Roles and Use Cases}
\label{sec-roles-and-use-cases}

\begin{figure}
	\includegraphics[width=\textwidth]{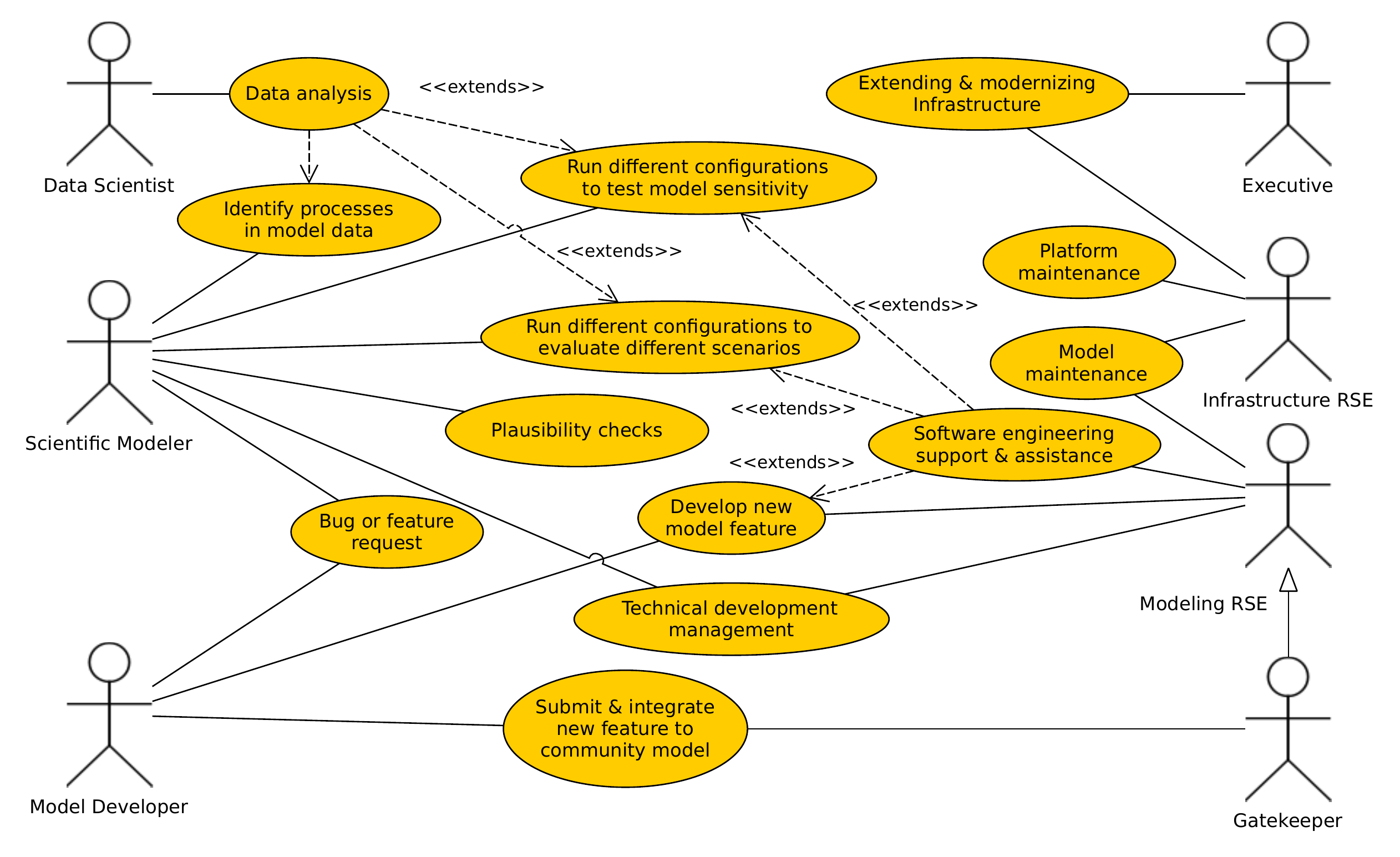}
	\caption{Roles and associated use cases in the ocean modeling domain.}
	\label{fig-roles-and-use-cases}
\end{figure}

Our analysis revealed seven distinct roles which interact in twelve specific use cases during modeling, maintenance, implementation and research, as illustrated via a UML use case diagram in Section~\ref{fig-roles-and-use-cases}.
These roles can be seen as specific perspectives and distinct kinds of interacting with ocean system models which can be performed by different actors.
In current practice of ocean modeling, individual scientists often take multiple of the identified roles.

The \emph{Data Scientist} performs analytics on data computed by models, produced by lab experiments, and obtained from observations in nature.
This role is especially involved in the data comparison process, discussed in Section~\ref{sec-data-comparison-process}, and contributes to model output analyses performed by the Scientific Modeler.

The \emph{Scientific Modeler} develops new and adapts or extends existing scientific models to address newly discovered findings.
Scientific Modelers are software users parameterizing and running the model to conduct research.
They also check the underlying theory and correct plausibility errors.
They contribute to the evaluation of models and their sensitivity to parameter changes, run specific scenarios, e.g., climate change scenarios, and provide insights into these scenarios.
For the scenarios, Scientific Modelers interact with the Data Scientist for data analysis.
Through their specific perspective as model users, Scientific Modelers issue bug reports and feature requests for the Infrastructure Research Software Engineer (RSE) (see below).

The \emph{Model Developer} is a software developer and responsible for transferring the scientific models into code and has to consider the scientific aspect of the code as well as technical aspects, like parallelization.
Model Developers also deploy the software, and submit feature or bug requests to the Infrastructure RSE.

\emph{Research Software Engineers~(RSE)} develop and maintain code, build systems, ensure software quality, and design software architectures.
However, there are two kinds of RSEs which we found in our analysis:
The \emph{Modeling RSE} is often a former scientist with extensive experience in the roles Scientific Modeler or Model Developer.
He or she supports the Model Developer, communicates best practice, guidelines and the envisioned architecture.
Modeling RSEs may also be responsible for maintenance and environment setup, like installation of development tools or configuration of the operating system.
The \emph{Infrastructure RSE} perform model and platform maintenance and advance the infrastructure layer.
They also provide technical support to Scientific Modelers and Model Developers, usually via bug fixing and implementing features. 
In these roles, RSE also coordinate joint efforts.

The \emph{Gatekeeper} is a special type of Modeling RSE and is responsible to control contributions to the community models, as established in open source software projects.\cite{shah_motivation_2006}
Gatekeeper interact with Model Developers and Modeling RSEs from various institutions to ensure quality standards in submissions and clean up code in collaboration with the contributors.
In contrast to Modeling RSEs who are often from the same organization, the Gatekeeper is from the software hosting organization and handles submissions form different research institutions.
Therefore, they are also software engineers, but are responsible for ensuring coding styles and best practices in the common code base and the upstream process.

The \emph{Executive} role decides on larger planning tasks, especially concerning the software infrastructure used by model developers, scientific modelers, and RSEs alike.
Among other tasks, they allocate resources and set priorities.

Table \ref{tbl-mapping-use-cases-processes} provides a survey of the relationship between the use cases from Figure~\ref{fig-roles-and-use-cases} and the processes / tasks that will be presented in Section~\ref{sec-main-process} to Section~\ref{sec-infrastructure-process}.
We employ BPMN for process modeling in the following subsections.
Table~\ref{tab:BPMNNotationElements} provides an overview of some BPMN notation elements that are used in this paper.

\clearpage

{
\begin{longtable}[!htb]{p{.2\textwidth} | p{.73\textwidth}}
	Use Case & Description \\
	\hline
	\hline
	\endhead
	Data analysis &
	Data analysis is performed in the \process{Data Comparison} process~(cf. Section~\ref{sec-data-comparison-process}). This and the \task{Data Analysis} task of the \process{Overall Ocean Science} process~(cf. Section~\ref{sec-main-process}) both cover data analysis.
	The former addresses in depth analyses of model and data with the aim to improve models, while the latter covers scenario and sensitivity study analyses.\\
	\hline
	Identify processes in model data &
	To identify emerging properties and processes utilizing the \task{Data Analysis} in \process{Overall Ocean Science} process~(cf. Section~\ref{sec-main-process}).\\
	\hline
	Run different configurations to test model sensitivity &
	Model sensitivity follows the \process{Deployment and Execution} sub-process for sensitivity studies).
	The analysis part of this use cases is covered by the \task{Data Analysis} of the \process{Overall Ocean Science} process~(cf. Section~\ref{sec-main-process}).\\
	\hline
	Run different configurations to evaluate different scenarios &
	Scenarios are used for in-depth analyses of specific real world scenarios, e.g., +2°C warming of the Earth.
	Here the \process{Deployment and Execution} sub-process with its scenario branch is utilized.
	The analysis part is covered by the \task{Data Analysis} task of the \process{Overall Ocean Science} process~(cf. Section~\ref{sec-main-process}).\\
	\hline
	Bug or feature request & During the use of models and development of new model extensions (e.g., following the \process{Modeling} process), new bugs in existing code, libraries and platform components may emerge, as well as, shortcomings of these artifacts. 
	Thus, Scientific Modelers and Model Developers may file bug and feature requests. \\
	\hline
	Plausibility Checks & Plausibility checks are used to evaluate whether a model corresponds with real world data and the expectations of Model Developers.
	This use case corresponds to the \task{Analyze Test Results} task of the \process{Test} sub-process.\\
	\hline
	Develop new model feature & 
	Incorporating new features into a model is mostly driven by the Model Developer, but can also be initiated by the Modeling RSE, depending on the nature of the addition.
	This follows the \process{Modeling} process~(cf. Section~\ref{sec-modeling-process}) and in particular the \process{Model Development} and \process{Test} sub-process.\\
	\hline
	Technical development management & Managing and communicating the architecture, of other technical/non-scientific aspects and new features.
	This is mainly driven by the Modeling RSE and can be initiated by the Scientific Modeler. \\
	\hline
	Submit \& integrate new feature in community model &
	Features and bug fixes created by Model Developer or Modeling RSE
	which are deemed relevant for the community are submitted to the Gatekeeper and integrated into the community model following the \process{Upstream Model} process~(cf. Section~\ref{sec-upstream-model}).\\
	\hline
	Bugs or feature request & Scientific Modelers and Model Developers may file bug and feature requests which are then addressed by the Infrastructure RSE.\\
	\hline
	Extending and modernizing infrastructure &
	Infrastructure RSE perform infrastructure modernization following the \process{Infrastructure} process~(cf. Section~\ref{sec-infrastructure-process}).
	The Executive role is involved to decide whether a specific effort is performed.\\
	\hline
	Model maintenance &
	While Model Developers and Scientific Modelers file bug reports and feature requests, the tickets are addressed by Infrastructure and Modeling RSEs following the \process{Maintenance} process~(cf. Section~\ref{sec-maintenance-process}).\\
	\hline
	Platform maintenance &
	While Model Developers and Scientific Modelers file bug reports and feature requests, the tickets are addressed by Infrastructure RSEs following the \process{Maintenance} process~(cf. Section~\ref{sec-maintenance-process}).\\
	\hline
	Software engineering support and assistance &
	This use case is an auxiliary use case extending many use cases to support the primary actor.
	For example, the Model Developer performs a plausibility check on a model feature and the Modeling RSE supports this effort by providing assistance regarding, e.g., configuration, architecture, setup, and community standards.\\
	\caption{Relationship between the use cases from Figure~\ref{fig-roles-and-use-cases} and the processes / tasks that will be presented in Section~\ref{sec-main-process} to Section~\ref{sec-infrastructure-process}.}
	\label{tbl-mapping-use-cases-processes}
\end{longtable}}

\newcommand{\img}[1]{\raisebox{-\totalheight}{\includegraphics[scale=0.6]{#1}}}
\begin{table}[!htb]
\begin{center}
\footnotesize
\begin{tabular}[t]{cp{.24\textwidth}|cp{.24\textwidth}} 
\img{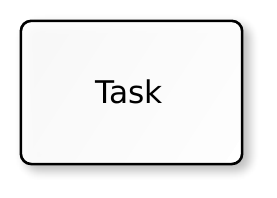} & 
A single task to be performed by a human or a computer. &
\img{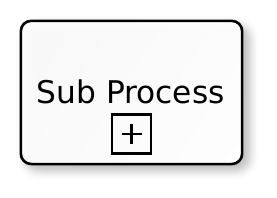} & 
Sub processes represent separate models for certain tasks or processes within a process, specified in an extra process model. \\
\hline
\img{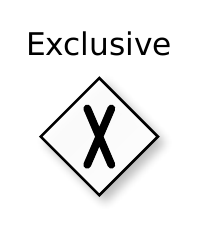} & 
The exclusive gateway changes the flow in the processes in one of the alternative directions. &
\img{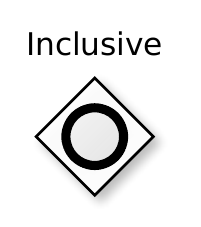} & 
The inclusive gateway allows to follow both path. \\
\hline
\img{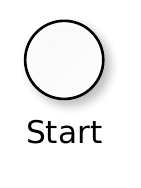} & 
The start event indicates where the process starts and which event triggers it. &
\img{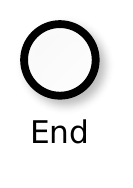} & 
The end event indicates the end of a process. \\
\hline
\img{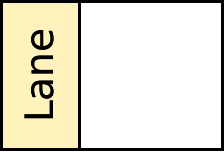} & 
\multicolumn{3}{p{0.6\textwidth}}{Lanes are used to describe that certain tasks or sub-processes of a process are performed by a specific role.} \\
\end{tabular}
\end{center}
\caption{BPMN elements employed in this paper.}
\label{tab:BPMNNotationElements}
\end{table}

\FloatBarrier

\subsection{The Overall Ocean Science Process}
\label{sec-main-process}

\begin{figure}[!htb]
	\includegraphics[scale=\imgscale]{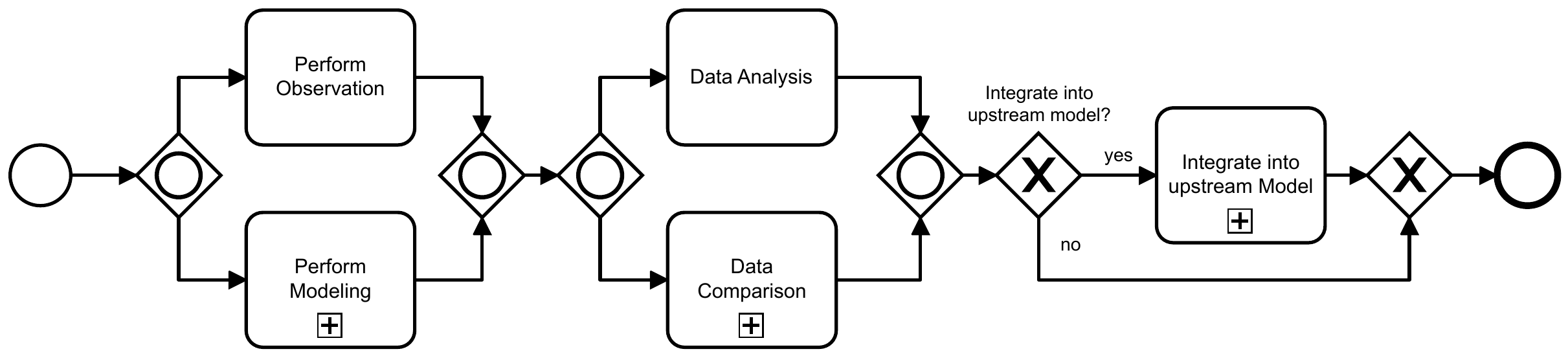}
	\caption{\process{Overall Ocean Science} process in ocean modeling.}
	\label{fig-main-process}
\end{figure}

The \process{Overall Ocean Science} process covers the work of ocean scientists and engineers on a top level, as shown in Figure~\ref{fig-main-process}.
In ocean science, the primary research activities are ocean observation and ocean modeling, followed by ocean data analysis and comparison. 
With the BPMN, we model activities, which may be single tasks or refined sub-processes. Table~\ref{tab:BPMNNotationElements} provides an overview of the BPMN elements employed in this paper.
In Figure~\ref{fig-main-process}, \task{Observation} and \task{Data Analysis} are modeled as tasks, which are not further refined in the present paper that focuses on ocean modeling.
We refined the remainder of the process \process{Overall Ocean Science} in three sub-processes indicated by $\vcenter{\hbox{\includegraphics[scale=.5]{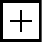}}}$ within the activity box.

Collecting data from the ocean is represented by the \task{Observation} task.
Ocean observations are related to ocean modeling, in particular as model input parameters and for model validation.
The \process{Modeling} process is one core activity in working with ocean system models covering model development, testing and model execution.
Details are covered in Section~\ref{sec-modeling-process}.
It can be run in parallel or independent of the \task{Observation} task (inclusive BPMN gateway).
Both, \task{Observation} and \process{Modeling}, produce data output which can then be analyzed independently, i.e., \task{Data Analysis}, or compared with other data sets in the \process{Data Comparison} process.
The \task{Data Analysis} task represents a wide range of activities dependent on the goal of the data analysis.
For example, ocean and earth system models are run for specific scenarios producing extensive model outputs, containing information on chemical, biological and physical processes.
These output datasets from the \process{Modeling} process are stored and analyzed by Data Scientists.
They may be interested in whether specific effects can be seen in the model data regarding their research questions. 
In case they identify new behavior or real world processes within the data, this may steer future observations and experiments or new analysis of existing observation data to test whether the findings have any real world correspondence. 
The \process{Data Comparison} process is a specific kind of data analysis where model output and observation data are compared to identify and to analyze differences for guiding observations and ocean modeling.
The process is discussed in more detail in Section~\ref{sec-data-comparison-process}.
Finally, the \process{Integrate into Upstream Model} process is an optional process addressing the integration of new model features into the upstream model, i.e., the research group or the institute that provides the community model.
As Figure~\ref{fig-main-process} shows, this may occur after a \task{Data Analysis} or a \process{Data Comparison}.
New contributions created in \process{Modeling} activities have the potential to advance the scientific community via additions to the upstream models.

\FloatBarrier
\subsection{The Modeling Process}
\label{sec-modeling-process}

\begin{figure}[!htb]
	\includegraphics[scale=\imgscale]{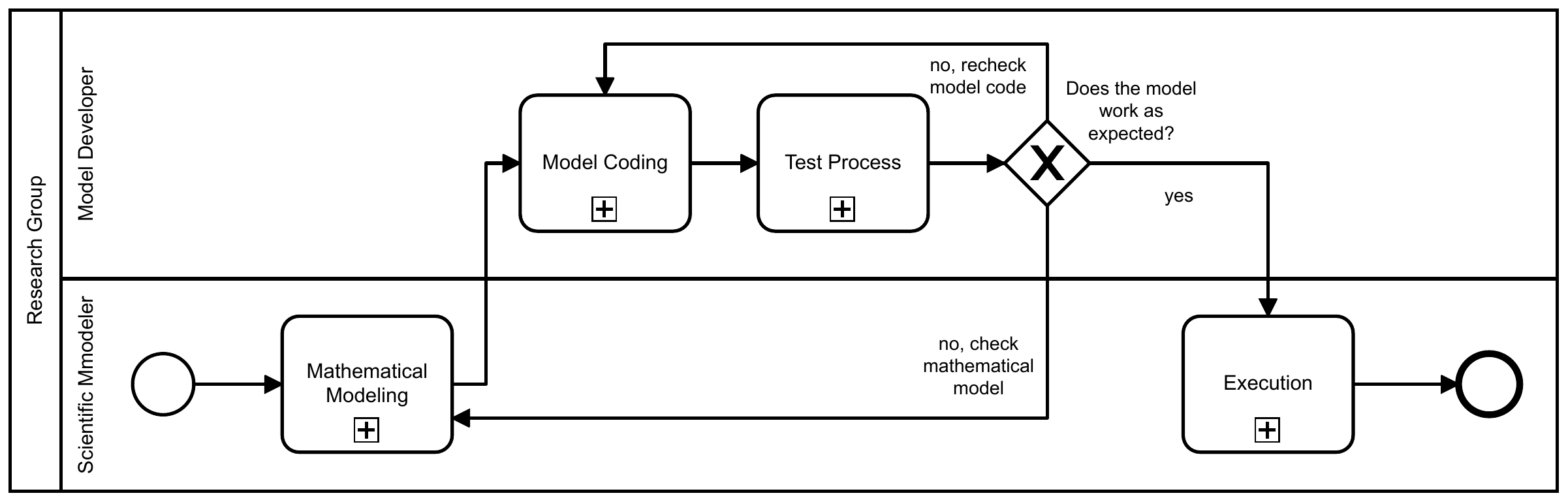}
	\caption{Overview chart of the \process{Modeling} process.}
	\label{fig-modeling-process}
\end{figure}

The \process{Modeling} process involves Scientific Modelers and Model Developers, as illustrated in Figure~\ref{fig-modeling-process}.
At its core, this process is a sequence of creating, adapting and modifying new or existing models.
Followed by testing, configuration, (remote) deployment, and execution.
Not all phases are always present in every modeling project or experiment.

Scientists involved in this process are in most cases Model Developers of a small part of the model and Scientific Modelers for the rest.
There is a wide range of triggers to engage in modeling including new hypotheses, research questions, or new observation data.
Again, we refined this process into three sub-processes indicated by $\vcenter{\hbox{\includegraphics[scale=.5]{figures/bpmn-boxplus}}}$ within the activity box (see Figure~\ref{fig-modeling-process}). \\[3mm]

\begin{figure}[!htb]
	\includegraphics[scale=\imgscale]{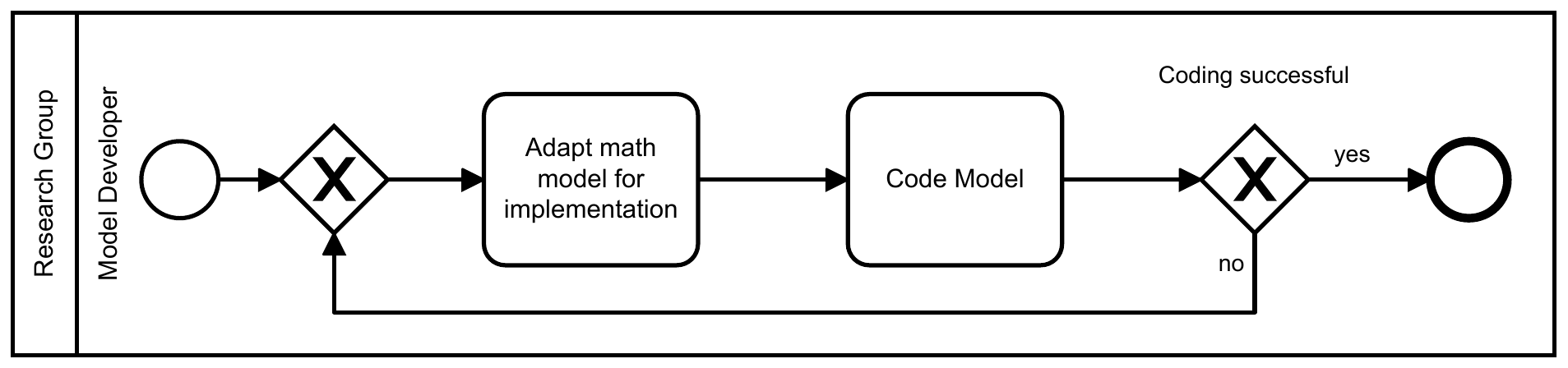}
\caption{\process{Model Development} sub-process}
\label{fig-model-development-process}
\end{figure}

\paragraph{Model Development sub-process}
\label{p-model-development}
This sub-process, depicted in Figure~\ref{fig-model-development-process}, includes mathematical model adaption and coding of the model.
Especially, the first task can be performed with pencil and paper, including manual equation solving.
This includes adapting the model to other resolutions in space and time that may require analytical integration.
In some cases this is checked using a numerical solver.
This is done, for instance, with tools such as Matlab and Octave\cite{quarteroni2010scientific} or libraries like Ferret\cite{Ferret2020} which can be used as an embedded domain-specific language in Python.

\begin{figure}[!htb]
	\includegraphics[scale=\imgscale]{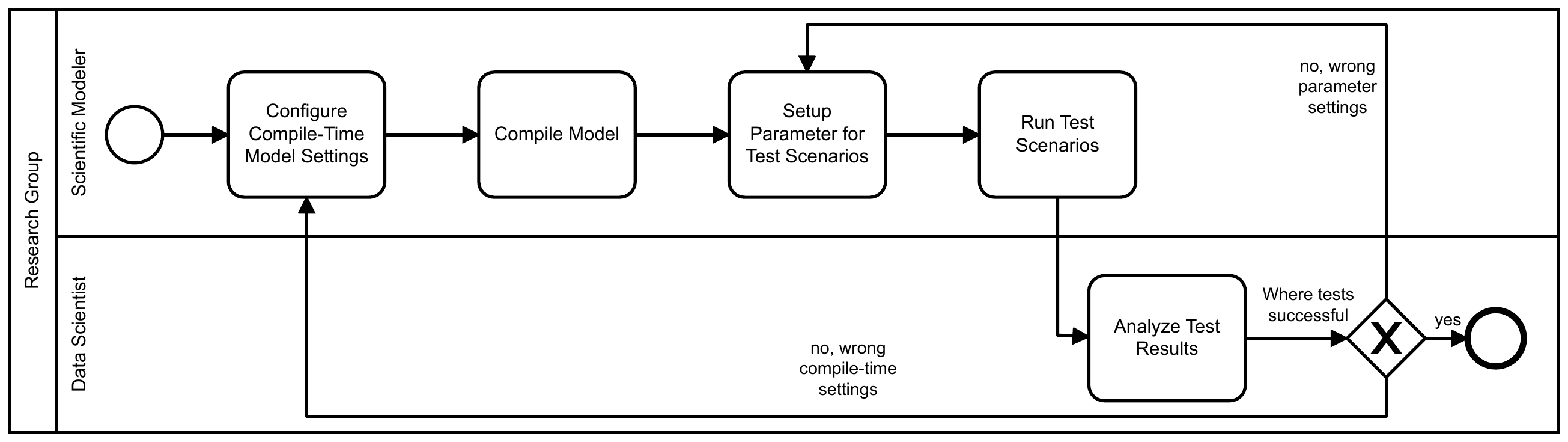}
	\caption{\process{Test} sub-process.}
	\label{fig-test-process}
\end{figure}

\paragraph{Test sub-process}
\label{p-test-process}
This sub-process tests whether the model behaves as expected and checks the test results for plausibility, as it is illustrated in Figure~\ref{fig-test-process}.
This can be done on a HPC or local machine depending on input data, runtime and computation power.
Usually, a model is tested with predefined setups against existing model results produced previously with the same setup.
Depending on the model, it is sometimes possible to test only parts by deactivating code through preprocessor flags.
In particular, when a new model component has been introduced, it is deactivated via flags and the model must still produce the same output as before the contribution was added.

As shown in Figure~\ref{fig-test-process} there are two tasks regarding model configuration.
The first configures code based on preprocessor flags or other compile time settings, which influence the compilation and code generation.
After compilation, additional parameters are configured via input files, with Fortran usually via namelists.

The plausibility checks are performed based on previous model runs.
Thus, the experiment setup and input data is reused for the tests.
As discussed above, one check is whether the model produces still the same output as before when the contribution is deactivated.
Here model outputs can be compared automatically to known reference data.
The specific model runs for these checks are scripted and automated.
Metamorphic testing \cite{MetamorphicTesting2016} can be utilized to test unknown results if a transformation of previous results is possible.
In ocean system modeling, such (metamorphic) transformations / relation are usually available.\cite{Hiremath2021}

However, with some models on some hardware setups, the output may still differ for numerical reasons or concurrency anomalies.
In these cases discrepancies must be analyzed manually. 
Subsequent tests include the new contributions where changes are expected but the scientific justification must be discussed and, if necessary, the reference data should be updated.
Thus, simple matching of outputs does not suffice to test the model.
Tests must also be operated manually e.g. with special visualizations in post-processing to validate the results.
Parameter fixes are performed within the \process{Test} sub-process, while potential programming errors are addressed by invoking the \process{Model Development} sub-process, as was depicted in Figure~\ref{fig-model-development-process}.

\begin{figure}[htb]
	\includegraphics[scale=\imgscale]{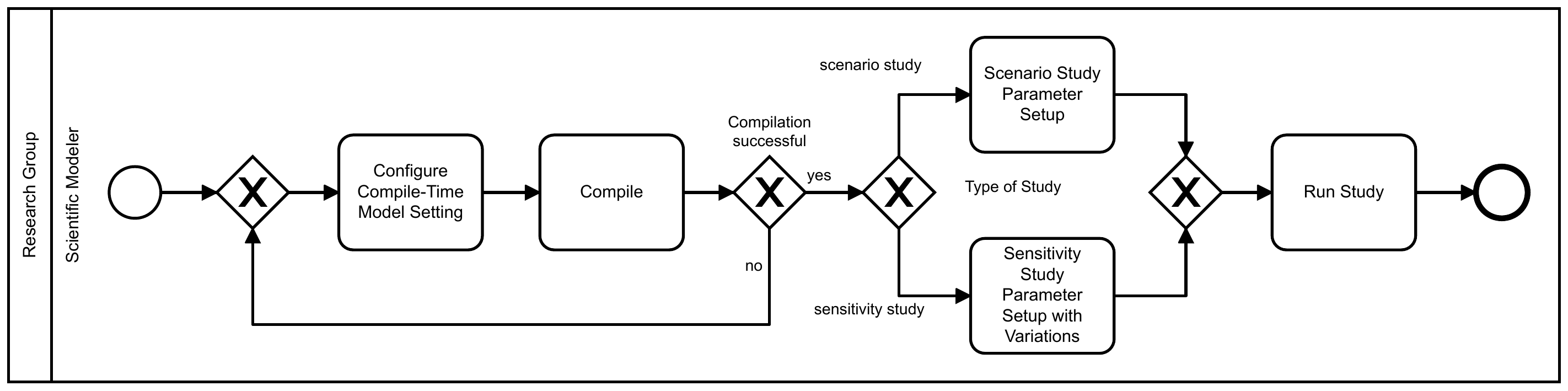}
	\caption{\process{Deployment and Execution} sub-process.}
	\label{fig-model-run-process}
\end{figure}

\paragraph{Deployment and Execution sub-process}
\label{p-deployment-and-execution}
The Deployment and Execution sub-process~(cf.~Figure~\ref{fig-model-run-process}) resembles the testing process to some extent, as both start with \task{Configure Compile-Time Model Settings}, followed by \task{Compile}, a parameter setup task, and the execution of the models.
Thus, the configuration and experiment setup may be reused to a large extent depending on the requirements of the experiment.
In most cases the testing and the execution are performed on the same platform.
However, some early tests might have been run on a workstation.
This highly depends on the model and its ability to produce identical or near to identical results on different platforms.

The configuration of the model is often changed in this sub-process to, for example, run models for a longer time to cover an extended period of time within the model.
Depending on the general aim of the experiments, a fixed setup is used, data from previous runs are fed into the model, or variations of configurations are used to test model sensitivity.
A \emph{Scenario Study} uses a predefined setup, optionally including model state~(data) from previous runs, and executes the model multiple times with the fixed setup.
A \emph{Sensitivity Study} varies parameters to estimate the sensitivity to specific parameter changes.
These different model setups are then run as separate experiments either in parallel or consecutive, depending on data dependencies and availability of computing resources. 

Once the runs are completed, the results are analyzed similarly to the tests from the plausibility checks.
In many cases specific scripts are used to post-process resulting data and analyze the content.
Scientists use tools, domain-specific languages, and scripts to process the model output and scrutinize the results.

\subsection{The Data Comparison Process}
\label{sec-data-comparison-process}

\begin{figure}[!htb]
	\includegraphics[scale=\imgscale]{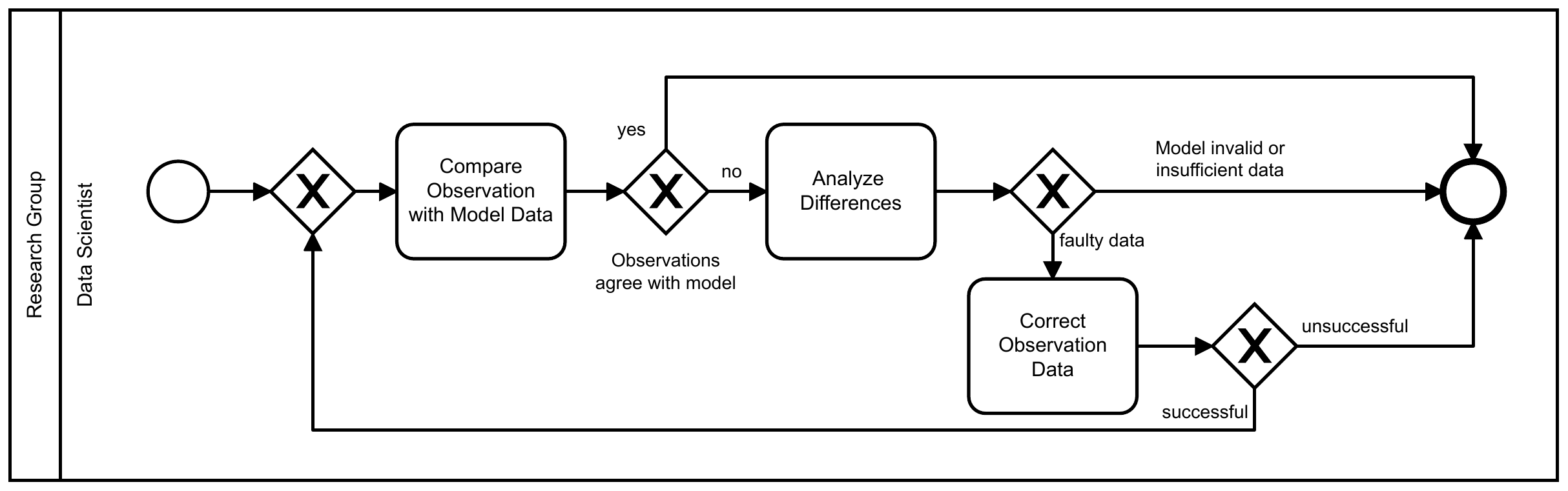}
	\caption{Overview chart of the \process{Data Comparison} process.}
	\label{fig-data-comparison-process}
\end{figure}

The \process{Data Comparison} process, depicted in Figure~\ref{fig-data-comparison-process}, compares observation data and model outputs.
This is done for various purposes.
The two major goals we found in the interviews were~(1) to validate the model,~and (2) to validate and describe the observation data via a model.

This process compares the datasets.
In case observations and the model agree, nothing more has to be done in this process.
However, in case differences arise, these have to be analyzed and reflected on in greater detail.
We identified three major outcomes of the analysis:
	\begin{enumerate}
		\item The difference stems from an invalid model.
		\item The data collected could be insufficient. Thus, additional data is needed which may require new observation campaigns.
		\item There could be faulty data, for instance caused by broken sensors. In some cases, this faulty data may be cleaned or adjusted. For good scientific practice, such cleansing has to be documented for proper data provenance.
	\end{enumerate}

\subsection{The Integrate into Upstream Model Process}
\label{sec-upstream-model}

\begin{figure}[!htb]
	\includegraphics[scale=\imgscale]{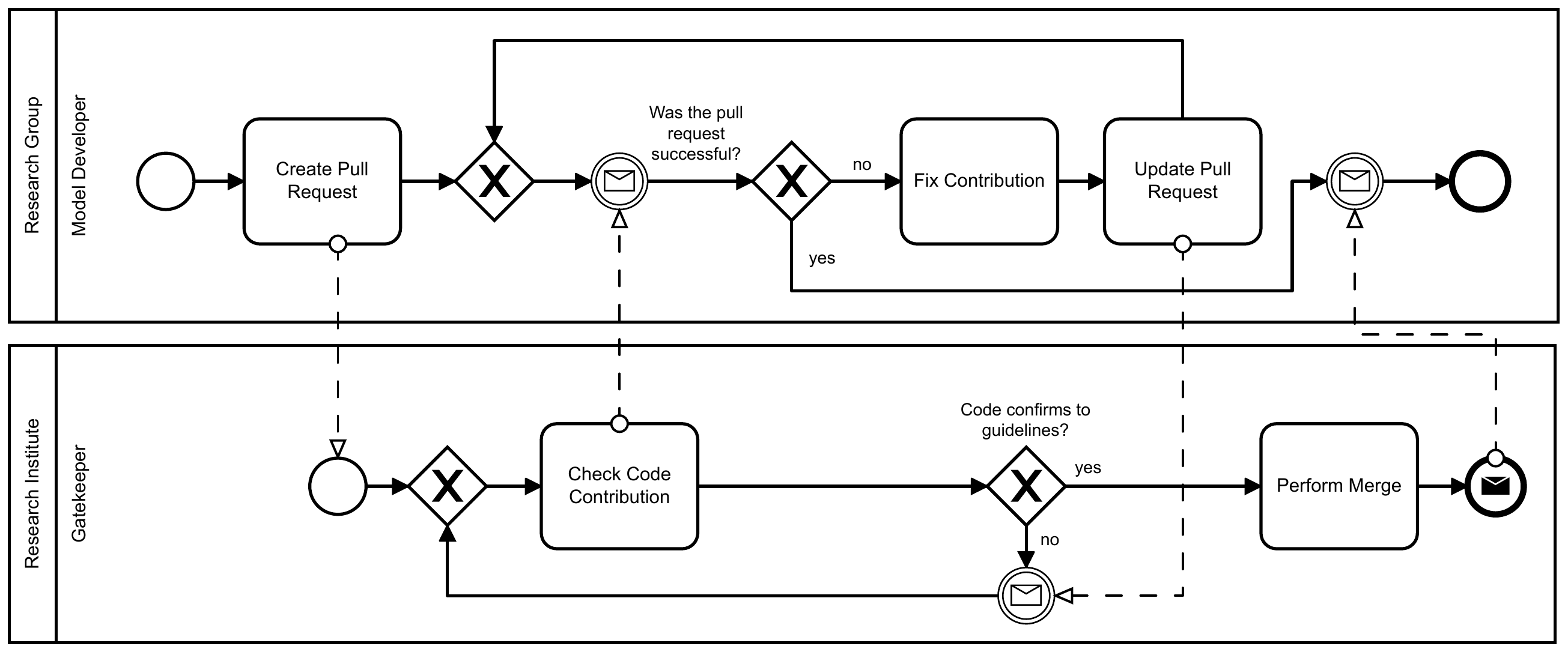}
	\caption{\process{Integrate into Upstream Model} process via pull requests.}
	\label{fig-model-pull-request-process}
\end{figure}

The process for integration of code into the upstream model is depicted in Figure~\ref{fig-model-pull-request-process}.
This process is not used in all modeling projects for various reasons.
Some projects do not have an upstream~(parent or community) model code base.
Sometimes patches or tarballs are sent to other groups as an informal approach to sharing contributions.
In other cases there is an upstream model but no technical~(e.g. Git repository) and social~(e.g. model maintainer or RSE) facility to handle revisions sent to the community model hosting organization.

The process involves a Model Developer and a Gatekeeper, as depicted in Figure~\ref{fig-model-pull-request-process}.
Essentially, a Model Developer issues a pull request. 
In case GitHub or GitLab is used, this is done using the corresponding functionality.
Other cases use e.g., tarballs or patch files.
The pull request is sent to the Gatekeeper to review and test the code.
Here, the Gatekeeper can rely on the tests from the \process{Test} sub-process.
In case the code does not pass the tests, the Gatekeeper can modify the code, or make change requests.
The contribution is handed back to the Model Developer to confirm the changes and fix issues the Gatekeeper pointed out.
Thereby some quality checks can be performed by the Gatekeeper.
In case the Gatekeeper accepts the code, it is merged with the existing model code by the Gatekeeper and will become a part of the next upstream model release.

\subsection{The Maintenance Process}
\label{sec-maintenance-process}

\begin{figure}[!htb]
	\includegraphics[scale=\imgscale]{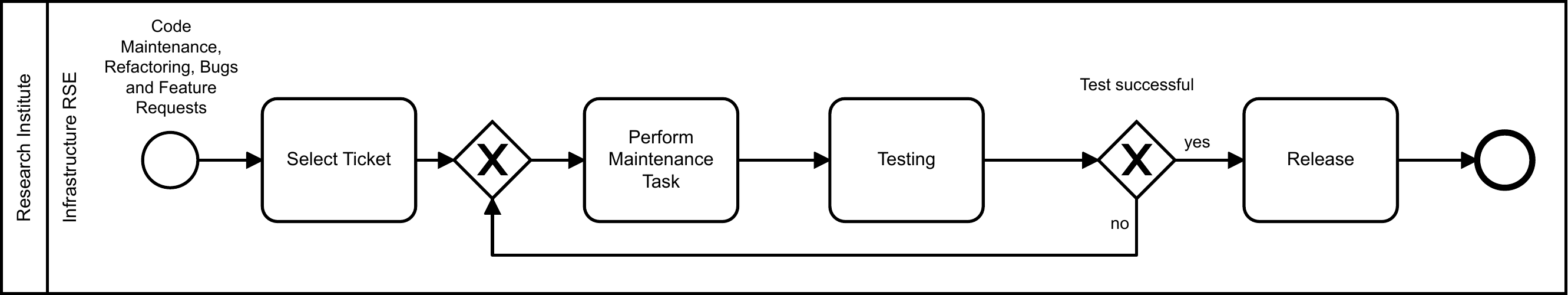}
	\caption{\process{Maintenance} Process.}
	\label{fig-maintenance-process}
\end{figure}

The previous processes have primarily addressed the work of Model Users, Model Developers, Modeling RSEs, and Gatekeepers.
The \process{Maintenance} process is primarily used by Infrastructure RSEs.
They may use ticket systems to coordinate their maintenance efforts.
Thus, each ticket will trigger the process.
However, not all institutions use ticket systems. 
Thus, the process may be started as a result of an email conversation or on Post-it notes on a Kanban board. 
This process, depicted in Figure~\ref{fig-maintenance-process}, is very similar to standard industrial software maintenance processes.

Essentially, the Infrastructure RSE selects the next ticket and starts working on it.
Subsequently, the maintenance task is performed and tested.
In case of infrastructure software, like IO library functions, it will be tested with unit tests which could be executed on the Infrastructure RSE's workstation.
Other tests, especially tests with models, are usually executed on the same hardware as the model itself.
This is due to limited modularization of the models and the influence of the hardware setup on the model, especially when testing parallelization.
In case the tests are successful, the changes are made available.
In community models, the code is released and formally introduced.
In some cases changes are only circulated within the institution.
The code may also be pushed to other repositories or sent via pull request to upstream code bases.
This is handled very differently by institutions, and community model projects.

\subsection{The Infrastructure Extension Process}
\label{sec-infrastructure-process}

\begin{figure}[!htb]
	\includegraphics[scale=0.55]{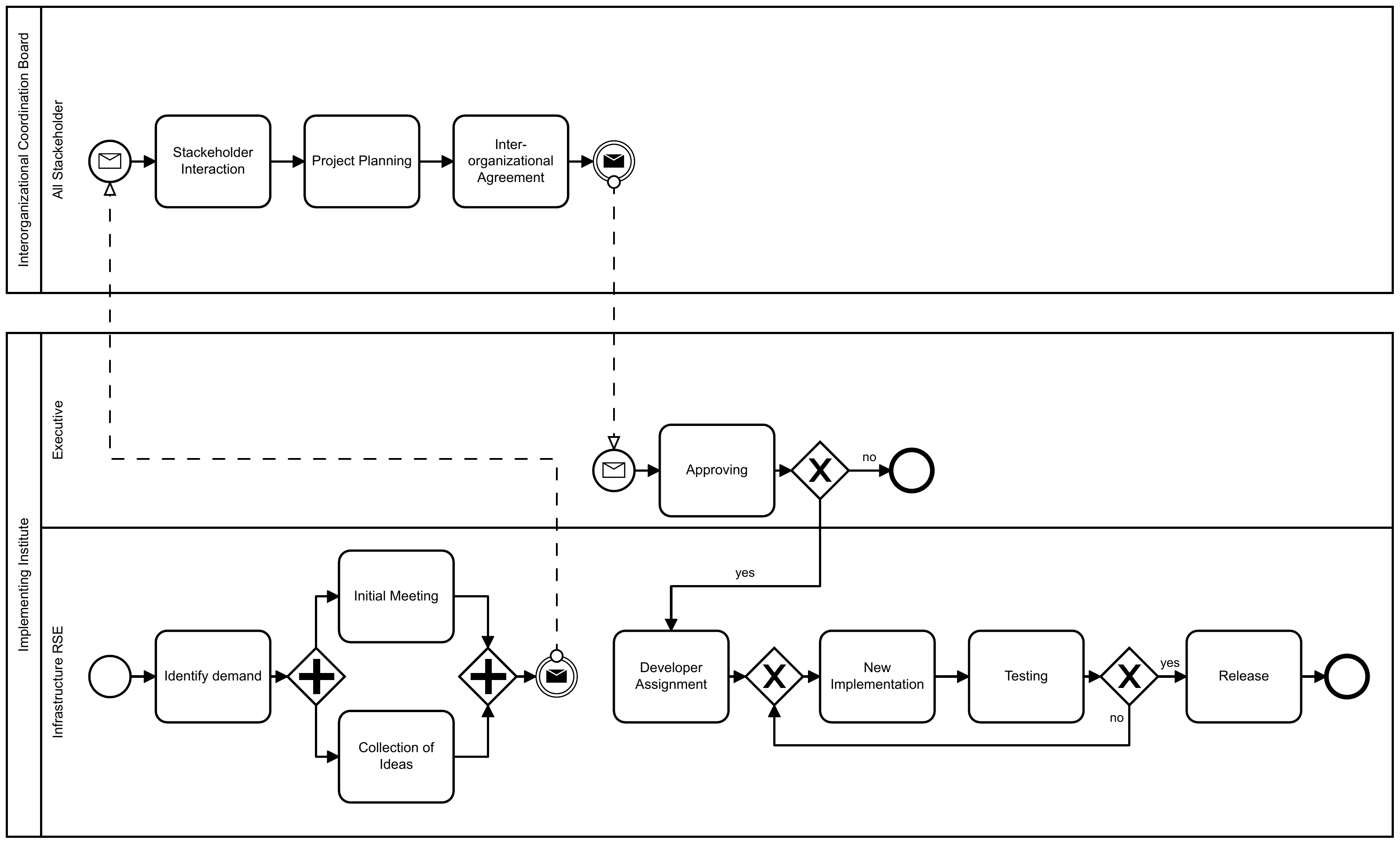}
	\caption{\process{Infrastructure} Process}
	\label{fig-infrastructure-extension-process}
\end{figure}

In case of larger changes to the infrastructure which affect multiple stakeholders or a significant share of the modeling project, the \process{Infrastructure Extension} process is followed.
Typical infrastructure changes are investments in hardware and system software. 
Due to the severity of such changes, all stakeholders should be involved in the process.
This is especially the case when multiple institutions are involved in using and/or maintaining the model.
This process, depicted in Figure~\ref{fig-infrastructure-extension-process}, is either triggered by Infrastructure RSEs or institutions themselves when a new demand is identified.
\noindent

These demands can be introduced, e.g., by new HPC hardware, additional partners, and new system software.
Based on the demand, initial meetings are held to collect ideas on how to address the new requirements.
This is usually a meeting of Infrastructure RSEs.
The next task is \task{Stakeholder Interaction}, i.e., scientists and personnel from other institutions covering RSEs of both kinds, Model Developers, and to some extent Scientific Modelers.
The aim is to finalize the approach and achieve a common understanding of the changes.
Furthermore, information is acquired regarding when the changes will be made available, as the changes might break scientific code or delay model runs.
Thus, all stakeholders should be involved in the planning of milestones and release dates.
Subsequently, a project plan is conceived and prepared for the Task \task{Interorganizational Agreement}.
Finally, the Executive has to authorize the project and allocate resources to it.
The remainder follows a typical development process, i.e., developers get their task assignments, the changes get implemented and tested, and finally released.

\section{Related Work}\label{sec-related-work}

Development of scientific software has been discussed in reports, case studies, surveys, and literature reviews.
In the following subsections, we discuss surveys, mapping studies, and case studies. 
To the best of our knowledge, there exist no related work that employs thematic analysis for scientific modeling, as we do.

\subsection{Surveys and Mapping Studies}

A survey on software development in the science and engineering domain by Eisty et al.\cite{Eisty:2019} focuses on gaining insights on the processes used in research software development.
While the paper does not dwell on the specific tasks in and around the development of scientific software, they discuss roles and types of development processes based on the responses from participants working on research software projects.
Most projects have only a few developers, while larger developers groups are rare.
In the survey, they defined five distinct roles called Developer, Architect, Manager, Executive and Other.
Unfortunately, they do not elaborate on the use cases or responsibilities of these roles.
Developers may correspond with our Scientific Modeler and Model Developer, while Architect is to some extent governed by Modeling RSE in our context.
In the area of processes, Eisty et al.\cite{Eisty:2019} show a preference for agile and ad-hoc process models, which corresponds with our findings by providing alternative routes in our process models.

Arvanitou et al.\cite{arvanitou_software_2021} describe the results of a systematic mapping study on the use of software engineering for scientific application development and their impact on software quality.
They describe a catalog of software engineering practices used in scientific software development and discuss the quality attributes of interest that drive the application of these practices. They survey software engineering practices per activity, such as design, implementation and testing. However, they do not survey roles as we do.

\subsection{Case Studies}

The FLASH project is a community project mainly supported by two institutions, but also external contributors.\cite{Dubey:2013}
Its domain is astrophysics and it has been developed over 20 years.
While they did not sketch specific workflows used in the development, the paper introduces two roles, i.e., software engineers and scientists, which are similar to our RSE and Scientific Modeler roles.
There seems to be no gatekeeping and the paper does not discuss branches as a means of development.

The second case study compared Scrum to a specific process for research and development called COMET-OCEP.\cite{Fonseca:2020}
They aim to develop a machine learning framework for person re-identification.
Initially, they applied Scrum and subsequently their new COMET-OCEP approach.
With the COMET-OCEP approach, they distinguish between two different roles, i.e. developer and researcher.
Their setup requires that these roles are represented by different persons to ensure that a dialog between the roles takes place, which would not be possible when one person takes multiple roles.
The case study resembles a new project, as it is build up from scratch utilizing a set of existing libraries used in machine learning.
The developer role is taken by trained developers who use established software engineering techniques, such as requirements specifications, UML diagrams and use cases.
Furthermore, they apply prototyping, unit testing and other standard development techniques.
All these software engineering methods and techniques show a positive impact on the case study, especially in context of the COMET-OCEP process.\cite{Fonseca:2020}
While these are promising results, our own interviews show that the premise of having separate representatives for the research and development related roles has no resemblance in the ocean modeling communities.
Typical ocean modeling projects work with large code bases which have grown for decades and might even have borrowed code from different sources following different guidelines.

The British Meteorologic Office Hadley Centre supports a large Unified Model which is developed by 100+ developers.\cite{Easterbrook:2009}
In an ethnographic study based on interviews and observations, they identified the development process, the culture and environment of the developers.
Similarly to our own findings, they categorized the involved personnel.
In contrast to roles, they use a layer concept to categorize duties and tasks.
They have a core IT team for the infrastructure and technical review, code owners managing features and handling new scientific code contributions, configuration managers, and contributing internal and external scientists.
The core development process at the Hadley Centre is lightweight, but much more strict than those we discovered in our interviews.
Essentially, they use 4 phases in their release process: implementation, fixing, testing, and releasing.
They use ticket systems~(Trac) to track features and bugs.
Code management is done with Subversion.
There is no discussion about using Git or utilizing a multi-branch approach.
Their development cycle starts with collecting and implementing tickets.
After 3 months they make a feature freeze and continue to fix issues within these features for another month.
Subsequently, they test the model by rerunning a set of experiments for another month.
Each contribution or change undergoes a two-step review process. 
First, the code owner (comparable to our Gatekeeper) performs a scientific review on the contribution.
Second, the IT team~(comparable to an infrastructure RSE) performs a second review addressing issues like code styles and performance.
Essentially, both perform some gatekeeping.
However, no process is sketched how the gatekeeping is realized.
For the reviews, they also employ a test harness which is a set of predefined configurations and experiments.
In comparison to our findings, there are many similar elements, like gatekeeping and roles, but the processes described by Easterbrook and Johns\cite{Easterbrook:2009} do not address work processes with greater detail.
They focused on the development process and the work culture in the institute and how scientists obtain a shared understanding of the model.
In contrast, we were more interested to identify processes used in different research groups and institutions to gain a broader overview.

\section{Conclusions and Future Work}\label{sec-conclusion}

Software development processes allow us to understand how scientists and engineers work and interact with each other in ocean system modeling.
This understanding is essential to provide support for all different roles and in consequence the scientists working on ocean system models.
We conducted semi-structured interviews and analyzed the interview results via thematic analysis, and employed process modeling for describing the elicited scientific software development processes in ocean science.

We have identified and modeled several distinct processes based on interviews with scientists and research software engineers.
They comprise a main process how scientists work with models and engage in modeling, a modeling process including testing and model execution, and a data analysis and comparison process.
For research software engineering, we identified processes to integrate new features into upstream models, perform maintenance, and work on the infrastructure for the ocean system models.
These processes show an ideal state and help to improve the software development and maintenance process.
As in practice the processes in the domain are not always used in such detail or are evolved from pragmatic decisions.
Existing processes can be reviewed with the help of the modeled processes or be analyzed with system engineering using interviews, TA and BPMN.

While the findings in this paper are of a qualitative nature, they provide a necessary understanding of how scientists and research software engineers work to allow to develop suitable methods and tools to improve their work.
Furthermore, we derived seven distinct roles within the domain of ocean modeling and how they interact to address typical use cases in modeling.
While scientists often take multiple roles, it is beneficial to understand the differences to be able to understand processes, refine them and subsequently improve the development environment, streamline methods, and provide automation and standardization.

In the future, we will provide an additional analysis on the characteristics and properties of the ocean modeling domain which provides insights on social interactions, tooling and collaboration.
Based on the findings of this paper, we will develop and evaluate tooling, especially domain-specific languages,\cite{ESE2017} to improve certain aspects of working with ocean system models.

Another area for future research is the comparison of our elicited scientific software development processes in ocean science with development processes in other areas of simulation-based system engineering, such as manufacturing or healthcare.

\section*{Acknowledgments}

Funded by the Deutsche Forschungsgemeinschaft (DFG -- German Research Foundation), grant no.\ HA 2038/8-1 -- 425916241.

\bibliographystyle{unsrturl}

\begin{thebibliography}{10}

\bibitem{Ruede2018}
Ulrich R{\"u}de, Karen Willcox, Lois~Curfman McInnes, and Hans~De Sterck.
\newblock Research and education in computational science and engineering.
\newblock {\em Siam Review}, 60(3):707--754, 2018.
\newblock \href {https://doi.org/10.1137/16M1096840}
  {\path{doi:10.1137/16M1096840}}.

\bibitem{CiSE2018}
Arne Johanson and Wilhem Hasselbring.
\newblock Software engineering for computational science: Past, present,
  future.
\newblock {\em Computing in Science {\&} Engineering}, 20(2):90--109, mar 2018.
\newblock \href {https://doi.org/10.1109/MCSE.2018.021651343}
  {\path{doi:10.1109/MCSE.2018.021651343}}.

\bibitem{Randell2018}
Brian Randell.
\newblock {Fifty Years of Software Engineering -- or --- The View from
  Garmisch}.
\newblock {\em CoRR}, 2018.
\newblock URL: \url{http://arxiv.org/abs/1805.02742}, \href
  {http://arxiv.org/abs/1805.02742} {\path{arXiv:1805.02742}}.

\bibitem{wanninkhof_global_2013}
Rik Wanninkhof, G-H Park, Taro Takahashi, Colm Sweeney, R~Feely, Yukihiro
  Nojiri, Nicolas Gruber, Scott~C Doney, Galen~A McKinley, Andrew Lenton,
  et~al.
\newblock Global ocean carbon uptake: magnitude, variability and trends.
\newblock {\em Biogeosciences}, 10(3):1983--2000, mar 2013.
\newblock \href {https://doi.org/10.5194/bg-10-1983-2013}
  {\path{doi:10.5194/bg-10-1983-2013}}.

\bibitem{ClimateModels2015}
K.~Alexander and S.~M. Easterbrook.
\newblock {The software architecture of climate models: a graphical comparison
  of CMIP5 and EMICAR5 configurations}.
\newblock {\em Geoscientific Model Development}, 8(4):1221--1232, 2015.
\newblock \href {https://doi.org/10.5194/gmd-8-1221-2015}
  {\path{doi:10.5194/gmd-8-1221-2015}}.

\bibitem{Braun2006}
Virginia Braun and Victoria Clarke.
\newblock Using thematic analysis in psychology.
\newblock {\em Qualitative Research in Psychology}, 3(2):77--101, 2006.
\newblock \href {https://doi.org/10.1191/1478088706qp063oa}
  {\path{doi:10.1191/1478088706qp063oa}}.

\bibitem{jensen_model-based_2009}
Ekkart Kindler.
\newblock Model-{Based} {Software} {Engineering} and {Process}-{Aware}
  {Information} {Systems}.
\newblock In {\em Transactions on {Petri} {Nets} and {Other} {Models} of
  {Concurrency} {II}}, pages 27--45, Berlin, Heidelberg, 2009. Springer.
\newblock \href {https://doi.org/10.1007/978-3-642-00899-3_2}
  {\path{doi:10.1007/978-3-642-00899-3_2}}.

\bibitem{SA2018}
Wilhelm Hasselbring.
\newblock Software architecture: Past, present, future.
\newblock In Volker Gruhn and R{\"u}diger Striemer, editors, {\em The Essence
  of Software Engineering}, pages 169--184, Cham, 2018. Springer.
\newblock \href {https://doi.org/10.1007/978-3-319-73897-0_10}
  {\path{doi:10.1007/978-3-319-73897-0_10}}.

\bibitem{chinosi_bpmn2012}
Michele Chinosi and Alberto Trombetta.
\newblock {BPMN}: {An} introduction to the standard.
\newblock {\em Computer Standards \& Interfaces}, 34(1):124--134, jan 2012.
\newblock \href {https://doi.org/10.1016/j.csi.2011.06.002}
  {\path{doi:10.1016/j.csi.2011.06.002}}.

\bibitem{seidman_interviewing_2019}
Irving Seidman.
\newblock {\em Interviewing as qualitative research: a guide for researchers in
  education and the social sciences}.
\newblock Teachers College Press, New York, 5th edition, 2019.

\bibitem{Kaiser2014}
Robert Kaiser.
\newblock {\em Qualitative Experteninterviews}.
\newblock Springer VS, Wiesbaden, 2014.

\bibitem{milde-koehn-18-german-asr}
Benjamin Milde and Arne K{\"o}hn.
\newblock Open source automatic speech recognition for {German}.
\newblock In {\em Speech Communication; 13th ITG-Symposium}, pages 1--5. VDE,
  2018.

\bibitem{ocenaudio}
{OcenAudio Development Team}.
\newblock ocenaudio: Easy, fast and powerful audio editor.
\newblock \url{https://www.ocenaudio.com/}, version 3.7.11, 2020.

\bibitem{RQDA}
Ronggui Huang.
\newblock Rqda: R-based qualitative data analysis.
\newblock \url{http://rqda.r-forge.r-project.org}, 2018.
\newblock {R Package Version 0.3-1}.

\bibitem{Stol2016}
K.~{Stol}, P.~{Ralph}, and B.~{Fitzgerald}.
\newblock Grounded theory in software engineering research: A critical review
  and guidelines.
\newblock In {\em IEEE/ACM 38th International Conference on Software
  Engineering (ICSE)}, pages 120--131, 2016.
\newblock \href {https://doi.org/10.1145/2884781.2884833}
  {\path{doi:10.1145/2884781.2884833}}.

\bibitem{shah_motivation_2006}
Sonali~K. Shah.
\newblock Motivation, {Governance}, and the {Viability} of {Hybrid} {Forms} in
  {Open} {Source} {Software} {Development}.
\newblock {\em Management Science}, 52(7):1000--1014, jul 2006.
\newblock \href {https://doi.org/10.1287/mnsc.1060.0553}
  {\path{doi:10.1287/mnsc.1060.0553}}.

\bibitem{quarteroni2010scientific}
Alfio Quarteroni, Fausto Saleri, and Paola Gervasio.
\newblock {\em Scientific computing with {MATLAB and Octave}}.
\newblock Springer, Berlin Heidelberg, 3rd edition, 2010.

\bibitem{Ferret2020}
{National Oceanic and Atmospheric Administration}.
\newblock Ferret -- an analysis tool for gridded and non-gridded data.
\newblock \url{https://ferret.pmel.noaa.gov/Ferret}, 2020.

\bibitem{MetamorphicTesting2016}
S.~Segura, G.~Fraser, A.~B. Sanchez, and A.~Ruiz-Cort{\'e}s.
\newblock A survey on metamorphic testing.
\newblock {\em IEEE Transactions on Software Engineering}, 42(9):805--824,
  2016.
\newblock \href {https://doi.org/10.1109/TSE.2016.2532875}
  {\path{doi:10.1109/TSE.2016.2532875}}.

\bibitem{Hiremath2021}
Dilip~J Hiremath, Martin Claus, Wilhelm Hasselbring, and Willi Rath.
\newblock Towards automated metamorphic test identification for ocean system
  models.
\newblock In {\em 2021 {IEEE}/{ACM} 6th International Workshop on Metamorphic
  Testing ({MET})}, pages 42--46. {IEEE}, jun 2021.
\newblock \href {https://doi.org/10.1109/met52542.2021.00014}
  {\path{doi:10.1109/met52542.2021.00014}}.

\bibitem{Eisty:2019}
Nasir~U. Eisty, George~K. Thiruvathukal, and Jeffrey~C. Carver.
\newblock Use of software process in research software development: A survey.
\newblock In {\em Proceedings of the Evaluation and Assessment on Software
  Engineering}, pages 276--282. ACM, 2019.
\newblock \href {https://doi.org/10.1145/3319008.3319351}
  {\path{doi:10.1145/3319008.3319351}}.

\bibitem{arvanitou_software_2021}
Elvira-Maria Arvanitou, Apostolos Ampatzoglou, Alexander Chatzigeorgiou, and
  Jeffrey~C. Carver.
\newblock Software engineering practices for scientific software development:
  {A} systematic mapping study.
\newblock {\em Journal of Systems and Software}, 172, feb 2021.
\newblock \href {https://doi.org/10.1016/j.jss.2020.110848}
  {\path{doi:10.1016/j.jss.2020.110848}}.

\bibitem{Dubey:2013}
A.~{Dubey}, K.~{Antypas}, A.~{Calder}, B.~{Fryxell}, D.~{Lamb}, P.~{Ricker},
  L.~{Reid}, K.~{Riley}, R.~{Rosner}, A.~{Siegel}, F.~{Timmes},
  N.~{Vladimirova}, and K.~{Weide}.
\newblock The software development process of {FLASH}, a multiphysics
  simulation code.
\newblock In {\em 5th International Workshop on Software Engineering for
  Computational Science and Engineering (SE-CSE)}, pages 1--8, 2013.
\newblock \href {https://doi.org/10.1109/SECSE.2013.6615093}
  {\path{doi:10.1109/SECSE.2013.6615093}}.

\bibitem{Fonseca:2020}
Jes{\'u}s Fonseca, Miguel De-la Torre, Salvador Cervantes, Eric Granger, and
  Jezreel Mejia.
\newblock Comet-ocep: A software process for research and development.
\newblock In Jezreel Mejia, Mirna Mu{\~{n}}oz, {\'A}lvaro Rocha, and Yadira
  Qui{\~{n}}onez, editors, {\em New Perspectives in Software Engineering},
  pages 99--112, Cham, 2020. Springer International Publishing.
\newblock \href {https://doi.org/10.1007/978-3-030-63329-5_7}
  {\path{doi:10.1007/978-3-030-63329-5_7}}.

\bibitem{Easterbrook:2009}
S.~M. {Easterbrook} and T.~C. {Johns}.
\newblock Engineering the software for understanding climate change.
\newblock {\em Computing in Science Engineering}, 11(6):65--74, 2009.
\newblock \href {https://doi.org/10.1109/MCSE.2009.193}
  {\path{doi:10.1109/MCSE.2009.193}}.

\bibitem{ESE2017}
Arne Johanson and Wilhelm Hasselbring.
\newblock Effectiveness and efficiency of a domain-specific language for
  high-performance marine ecosystem simulation: a controlled experiment.
\newblock {\em Empirical Software Engineering}, 22(4):2206--2236, aug 2017.
\newblock \href {https://doi.org/10.1007/s10664-016-9483-z}
  {\path{doi:10.1007/s10664-016-9483-z}}.

\end{thebibliography}

\end{document}